\documentclass[10pt,prx,aps,twocolumn,english,superscriptaddress,citeautoscript,preprintnumbers,amsmath,amssymb,floatfix,footinbib]{revtex4-2}
\usepackage[breaklinks=true,colorlinks,citecolor=blue,linkcolor=blue,urlcolor=blue]{hyperref}
\usepackage{amsmath}
\usepackage{graphicx}
\usepackage{dcolumn}
\usepackage{float}
\usepackage[utf8]{inputenc}
\usepackage[T1]{fontenc}
\usepackage{color}

\usepackage{mathtools}

\usepackage[super]{nth}
\usepackage{soul}
\renewcommand{\AA}{ \text{Å}}
\usepackage{xcolor}

\begin{document}

\clearpage
    
\title{Type-I superconductivity in a quasi-2D topologically nontrivial YbBi$_2$}

\author{Karolina G\'ornicka}
\thanks{These authors contributed equally to this work.}
\email{gornickaka@ornl.gov}
\affiliation{Materials Science and Technology Division, Oak Ridge National Laboratory, Oak Ridge, Tennessee 37831, USA}
\affiliation{Faculty of Applied Physics and Mathematics, Gdansk University of Technology, Narutowicza 11/12, 80-233 Gdańsk, Poland}
\affiliation{Advanced Materials Center, Gdansk University of Technology, Narutowicza 11/12, 80-233 Gdańsk, Poland}
\author{Sudip Malick}
\thanks{These authors contributed equally to this work.}
\email{sudip.malick@pg.edu.pl}
\affiliation{Faculty of Applied Physics and Mathematics, Gdansk University of Technology, Narutowicza 11/12, 80-233 Gdańsk, Poland}
\affiliation{Advanced Materials Center, Gdansk University of Technology, Narutowicza 11/12, 80-233 Gdańsk, Poland}
\author{Joanna Bławat}
\affiliation{National High Magnetic Field Laboratory, Los Alamos National Laboratory, Los Alamos, NM, USA}
\author{Michał Modrzejewski}
\affiliation{AGH University of Krakow, Faculty of Physics and Applied Computer Science, Aleja Mickiewicza 30, 30-059 Krakow, Poland}
\author{Hanna Świątek}
\affiliation{Faculty of Applied Physics and Mathematics, Gdansk University of Technology, Narutowicza 11/12, 80-233 Gdańsk, Poland}
\affiliation{Advanced Materials Center, Gdansk University of Technology, Narutowicza 11/12, 80-233 Gdańsk, Poland}
\author{Michał J. Winiarski}
\affiliation{Faculty of Applied Physics and Mathematics, Gdansk University of Technology, Narutowicza 11/12, 80-233 Gdańsk, Poland}
\affiliation{Advanced Materials Center, Gdansk University of Technology, Narutowicza 11/12, 80-233 Gdańsk, Poland}
\author{Federico Mazzola}
\affiliation{Department of Physics and Astronomy ‘Galileo Galilei’, University of Padova, Padova, Italy}
\affiliation{CNR-Istituto di Struttura della Materia (CNR-ISM), Strada Statale 14, km 163.5, 34149 Trieste, Italy} 
\author{Ivana Vobornik}
\affiliation{CNR-IOM Istituto Officina dei Materiali, I-34139 Trieste, Italy}
\author{Chiara Bigi}
\affiliation{Synchrotron SOLEIL, L’Orme des Merisiers, D\'epartementale 128, F-91190 Saint-Aubin, France}
\author{Jacob Cook}
\affiliation{Materials Science and Technology Division, Oak Ridge National Laboratory, Oak Ridge, Tennessee 37831, USA}
\author{Brenden R. Ortiz}
\affiliation{Materials Science and Technology Division, Oak Ridge National Laboratory, Oak Ridge, Tennessee 37831, USA}
\author{Andrew F. May}
\affiliation{Materials Science and Technology Division, Oak Ridge National Laboratory, Oak Ridge, Tennessee 37831, USA}
\author{Andrzej P. Kądzielawa}
\affiliation{AGH University of Krakow, Faculty of Physics and Applied Computer Science, Aleja Mickiewicza 30, 30-059 Krakow, Poland}
\author{John Singleton}
\affiliation{National High Magnetic Field Laboratory, Los Alamos National Laboratory, Los Alamos, NM, USA}
\author{Bartlomiej Wiendlocha}
\email{wiendlocha@fis.agh.edu.pl}
\affiliation{AGH University of Krakow, Faculty of Physics and Applied Computer Science, Aleja Mickiewicza 30, 30-059 Krakow, Poland}
\author{Tomasz Klimczuk}
\email{tomasz.klimczuk@pg.edu.pl}
\affiliation{Faculty of Applied Physics and Mathematics, Gdansk University of Technology, Narutowicza 11/12, 80-233 Gdańsk, Poland}
\affiliation{Advanced Materials Center, Gdansk University of Technology, Narutowicza 11/12, 80-233 Gdańsk, Poland}

\begin{abstract}

Intrinsic superconductivity in stoichiometric materials with nontrivial electronic topology remains uncommon, limiting opportunities to investigate how these two phenomena coexist within a single electronic system. Here, we report bulk type-I superconductivity below $T_c$ $\sim$ 0.9~K  in  YbBi$_2$, a layered rare-earth compound with a nonsymmorphic crystal structure and a quasi-two-dimensional Fermi surface. Thermodynamic and transport measurements establish the superconducting ground state, while quantum oscillations reveal exceptionally light carriers, with a cyclotron mass as low as 0.07 $m_e$, and a nonzero Berry phase of approximately 0.82 $\pi$. The latter closely matches the calculated value of the corresponding Wilson phase 0.99 $\pi$ for the corresponding orbit near a symmetry-protected band degeneracy. ARPES measurements show good agreement with key features of the calculated electronic structure, providing complementary experimental constraints on the normal-state band structure. The combination of intrinsic type-I superconductivity, light quasi-two-dimensional carriers, and signatures of nontrivial electronic topology identifies  YbBi$_2$ as a distinct stoichiometric platform for investigating superconductivity in a topologically nontrivial electronic environment.

\end{abstract}

\maketitle
\section{Introduction}

Realizing superconductivity in materials with nontrivial electronic topology remains an important challenge in the search for unconventional quantum states \cite{PRL.100.096407, Wray2010, Xu2014, NadjPerge2014}. While there are many topological materials, only a few demonstrate intrinsic superconductivity without chemical substitution, external pressure, or proximity coupling \cite{Zhu_2019, Yin2022}. Rare-earth compounds offer a further dimension in the search through the presence of localized $f$ electrons and their coupling to the electronic structure. Thus identifying stoichiometric rare-earth materials that realize intrinsic superconductivity and nontrivial band topology can provide an opportunity to explore a comparatively less studied electronic regime.
In this context, bismuth-based compounds provide a natural materials platform for this search. Strong spin-orbit coupling, combined with low-dimensional structural motifs, can generate Dirac-like dispersions and nontrivial band topology \cite{annurev, Sharma2026}, while superconductivity occurs in several Bi-based systems, including $\beta$-PdBi$_2$~\cite{Sakano2015, Iwaya2017}, PtBi$_2$~\cite{Changdar2025, Schimmel2024} and CaBi$_2$~\cite{Winiarski2016, PhysRevB.104.245112}. In the case of Yb compounds, superconductivity is rare and has been mainly associated to strongly correlated systems like  $\beta$-YbAlB$_{4}$ \cite{Nakatsuji2008}, and YbPd$_{2}$Sn \cite{YbPd2Sn_1998, YbPd2Sn_2003}. More recently, YbSb$_2$ \cite{Sato1999, Zhao2012, Dhara2025, YbSb2_PRL}  has emerged as a contrasting example of a quasi-two-dimensional type-I superconductor with a Dirac nodal-line electronic structure, as well as evidence for time-reversal-symmetry breaking in the superconducting state. These discoveries open up a broader landscape in the case of Yb based superconductors and raise the question of how superconductivity evolves in the presence of nontrivial band topology.

YbBi$_2$ is a promising candidate to investigate this regime of matter. This material crystallizes in a nonsymmorphic orthorhombic ZrSi$_2$-type structure consisting of a stacking of square Bi layers and Yb-Bi slabs, forming a strongly anisotropic electronic structure \cite{OHARA2000752, YbBi2_2022}. Previous quantum-oscillation measurements and first-principles calculations revealed quasi-two-dimensional Fermi-surface sheets and predicted nontrivial electronic states related to the nonsymmorphic character of the crystal symmetry \cite{OHARA2000752, YbBi2_2022}. However, these studies were not performed in the sub-kelvin regime, leaving the ground state of this system unknown. Whether superconductivity emerges in this nontrivial electronic environment therefore remained an open question.

Here, we report on the observation of bulk type-I superconductivity in YbBi$_2$ below $T_c$ $\sim$ 0.9 K and find its coexistence with a quasi-two-dimensional electronic structure carrying signatures of nontrivial topology. Magnetization, electrical transport, and heat-capacity measurements all confirm the bulk superconducting state, while quantum oscillations reveal multiple Fermi-surface sheets and exceptionally light carriers with a cyclotron mass as low as 0.07 $m_e$. The light quasi-two-dimensional orbit has a nonzero Berry phase of approximately 0.82 $\pi$, in close agreement with the calculated value. ARPES measurements constrain the quasi-two-dimensional electronic structure and agree well with key features of the calculated band structure. Together, these complementary results establish YbBi$_2$ as a rare stoichiometric Yb-based material combining bulk type-I superconductivity, light quasi-two-dimensional carriers, and signatures of nontrivial electronic topology. This provides a distinct platform to explore how superconductivity develops in a rare-earth system with nontrivial band structure.

\section{Results and Discussions}
\begin{figure*}
    \includegraphics[width=17cm, keepaspectratio]{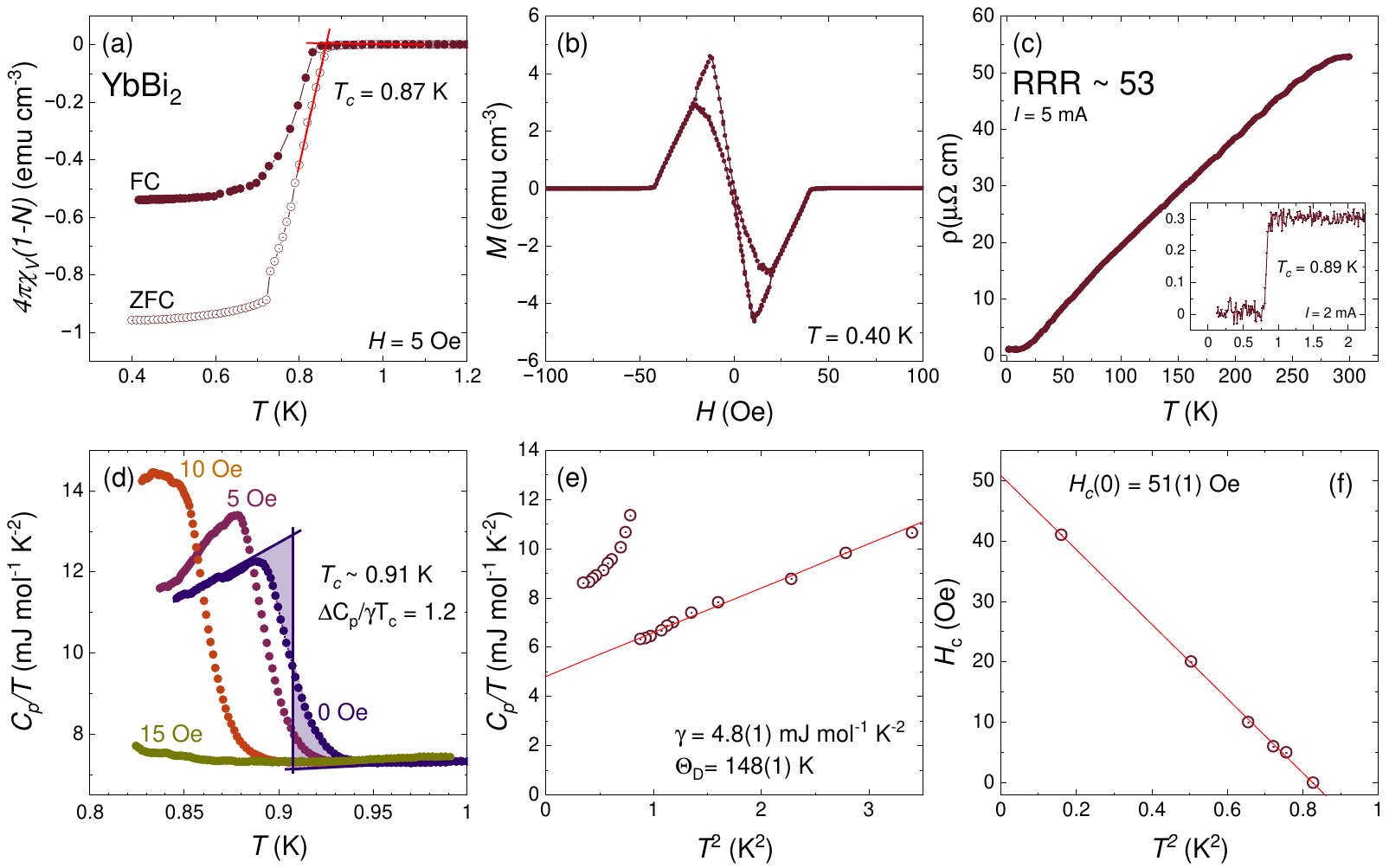}
	\caption{\label{YbBi2_sc} Superconducting properties of YbBi$_2$. (a)~Temperature-dependent zero-field-cooled (ZFC) and field-cooled (FC) magnetic susceptibility. (b)~Field-dependent magnetization measured at $T = 0.4$~K, indicating type-I superconductivity. (c)~Electrical resistivity over the full temperature range, with the superconducting transition shown in the inset. (d)~Heat-capacity measurements presented as $C_{\rm p}/T$ versus $T$ below 1~K in zero field and under low applied magnetic fields (5, 10, and 15~Oe). These measurements confirm the type-I superconductivity of YbBi$_2$. (e)~$C_{\rm p}/T$ versus $T^2$ together with the low-temperature fit described in the main text. (f)~Critical field ($H_{\rm c}$) versus $T^2$ with a linear fit used to determine the value of $H_{\rm c}(0)$. The data points were obtained from $M(H)$ measurements, except for the point at $H = 0$, which was determined from the zero-field heat-capacity measurement.}
\end{figure*}

\subsection{Crystallographic studies}

We begin by establishing the crystal structure, phase purity, and chemical composition of the grown YbBi$_2$ single crystals.  The room-temperature powder XRD pattern of crushed YbBi$_2$ single crystals together with the Le Bail refinement is presented in  Supplemental Material \cite{SM}. All major reflections can be indexed within the orthorhombic ZrSi$_2$-type structure (blue vertical bars), confirming that YbBi$_2$ crystallizes in the nonsymmorphic space group $Cmcm$, in agreement with previous reports~\cite{Maksudova1985}. The refined lattice parameters are: $a = 4.6176(7)~\AA$, $b = 17.014(5)~\AA$, $c = 4.414(2)~\AA$. No additional impurity phases were detected within the experimental resolution, apart from weak reflections from residual elemental Bi used as a flux. Note that the relative intensity of the Bi reflections increases over time. Consequently, Bi is considered to be one of the decomposition products of YbBi$_2$. The crystallographic orientation of selected plate-like single crystals was subsequently verified by XRD measurements performed on intact crystals. The crystallographic orientation of selected plate-like single crystals was checked by XRD measurements performed on intact crystals. The single-crystal XRD pattern (see Supplemental Material \cite{SM}) displays only three very narrow and split (K$\alpha1$/K$\alpha2$) reflections [0$k$0] at [$k$]~=~6, 10, and 12. Additional broad reflections originate from Bi, either as residual flux or as a decomposition product. Note that the crystal surface was exposed to air during the XRD measurements, which caused partial surface degradation.
The chemical composition was further examined using EDS analysis, which confirmed the expected Yb:Bi stoichiometry within experimental uncertainty and indicated no detectable elemental segregation. Together, powder XRD and EDS measurements demonstrate that the crystals are single-phase YbBi$_2$ with the faces oriented perpendicular to the [010] direction.

\subsection{Superconducting properties of YbBi$_2$}

Having established the structural and chemical integrity of the samples, we now turn to their magnetic, thermal, and transport properties. Previous studies were restricted to temperatures above 1.3~K, and therefore these investigations of YbBi$_2$ focused on its normal-state electronic properties, revealing large magnetoresistance and quantum oscillations at low temperatures~\cite{Ohara2000, YbBi2_2022}. By extending our measurements into the sub-kelvin regime, we found superconductivity in YbBi$_2$, as summarized in Fig.~\ref{YbBi2_sc}.

To characterize the superconducting transition, zero-field-cooled (ZFC) and field-cooled (FC) dc magnetic susceptibilities, defined as $\chi = M/H$, were measured upon warming under various small applied magnetic fields. Representative data collected at $H = 5$~Oe are shown in Fig.~\ref{YbBi2_sc}(a). After correcting for the demagnetization factor $N = 0.85$, determined from volume magnetization $M(H)$, a clear diamagnetic response is observed below $T_\mathrm{c} = 0.87$~K, marking the onset of superconductivity. The sharpness of the transition and the magnitude of the diamagnetic signal indicate a bulk superconducting state in YbBi$_2$. Further insight into the superconducting state is obtained from isothermal magnetization measurements. Fig. \ref{YbBi2_sc}(b) displays $M(H)$ measured at $T = 0.40$~K. At low fields, the magnetization initially varies linearly with field and then goes to zero near the critical field. This behavior is characteristic of a type-I superconductor~\cite{PhysRevB.72.212508, PhysRevB.85.174514, Zhao2012}. This is further confirmed by the heat capacity measurements below.

Transport measurements provide complementary evidence for superconductivity. The temperature dependence of the electrical resistivity $\rho(T)$, shown in Fig.~\ref{YbBi2_sc}(c), reveals a metallic behavior between 300~K and 1.8~K with a large residual resistivity ratio (RRR)~=~53, much larger than the previous report~\cite{YbBi2_2022} and indicative of high sample quality. The inset highlights the low-temperature region, where the resistivity drops sharply to zero at $T_{\mathrm{c}} = 0.89$~K, in good agreement with the transition temperatures determined from magnetic susceptibility measurements.

The final and most important test of the superconducting properties is provided by heat capacity measurements. The low-temperature heat capacity, plotted as $C_\mathrm{p}/T$ versus $T$ in Fig.~\ref{YbBi2_sc}(d), exhibits a pronounced peak near $T_\mathrm{c}$ in zero magnetic field. Applying an equal-entropy construction to enforce entropy balance between the normal and superconducting states yields $T_\mathrm{c} = 0.91~$K, consistent with the values obtained from magnetization and resistivity. The normalized heat-capacity jump $\Delta C/\gamma_\mathrm{n} T_\mathrm{c} = 1.2$ is slightly smaller than the weak-coupling BCS value of 1.43, suggesting weak electron-phonon coupling. Field-dependent heat-capacity measurements reveal further characteristics of the superconducting state. As shown in Fig.~\ref{YbBi2_sc}(d), the peak becomes sharper and more pronounced at $H = 5$~Oe compared to zero field, indicating a crossover from a second-order to a first-order phase transition. Such behavior is a hallmark of type-I superconductivity~\cite{Halperin1974, Zhao2012, Salis2021, Klimczuk2023} and is consistent with the magnetization results. To quantify the normal-state electronic properties, the heat-capacity data above $T_\mathrm{c}$ were analyzed using \begin{equation}
\frac{C_\mathrm{p}}{T} = \gamma + \beta T^{2},
\end{equation}
\noindent where $\gamma$ represents the electronic Sommerfeld coefficient, and the second term accounts for the phonon contribution. The fit, shown in Fig. \ref{YbBi2_sc}(e), yields $\gamma = 4.8 \text{ mJ} \text{ mol}^{-1} \text{K}^{-2}$  and a corresponding Debye temperature $\theta_\mathrm{D} = 148$ K calculated from the slope of the fitting line ($\beta$): \begin{equation}
\theta_\mathrm{D} = \left(\frac{12\pi^{4}Rn}{5\beta}\right)^{1/3},
\end{equation}
\noindent where $n = 3$ is the number of atoms per unit cell and $R$ is the gas constant. The Sommerfeld coefficient extracted from our low-temperature heat-capacity data is significantly smaller than the value reported previously at higher temperatures ($\gamma = 28\text{ mJ}\text{ mol}^{-1}\text{K}^{-2}$)~\cite{YbBi2_2022}. Having $\theta_\mathrm{D}$ and $T_\mathrm{c}$, the electron-phonon coupling constant can be calculated from the McMillan equation \cite{Mcmillan1968}:

\begin{equation}
\lambda_\mathrm{ep} = \frac{1.04+\mu^{*}\text{ln}(\frac{\theta_\mathrm{D}}{1.45T_\mathrm{c}})}{(1-0.62\mu^{*})\text{ln}(\frac{\theta_\mathrm{D}}{1.45T_\mathrm{c}})-1.04}  
\end{equation}

\noindent where the parameter $\mu^{*}$ is a Coulomb repulsion constant. Using a typical value $\mu^{*} = 0.13$, we obtained an electron-phonon coupling constant for YbBi$_{2}$ of $\lambda_\mathrm{ep} = 0.5$, indicative of a weak coupling superconductor. Finally, the superconducting phase boundary was mapped by extracting the critical field $H_\mathrm{c}$ as a function of temperature from magnetization measurements. The resulting $H_\mathrm{c}(T)$ curve, shown in Fig.~\ref{YbBi2_sc}(f), follows the expected quadratic behavior for a conventional type-I superconductor, yielding $H_\mathrm{c}(0) = 51(1)$~Oe and $T_\mathrm{c} = 0.91(1)$~K.

\subsection{Quantum Oscillations}
\begin{figure*}
	\includegraphics[width=17.5cm, keepaspectratio]{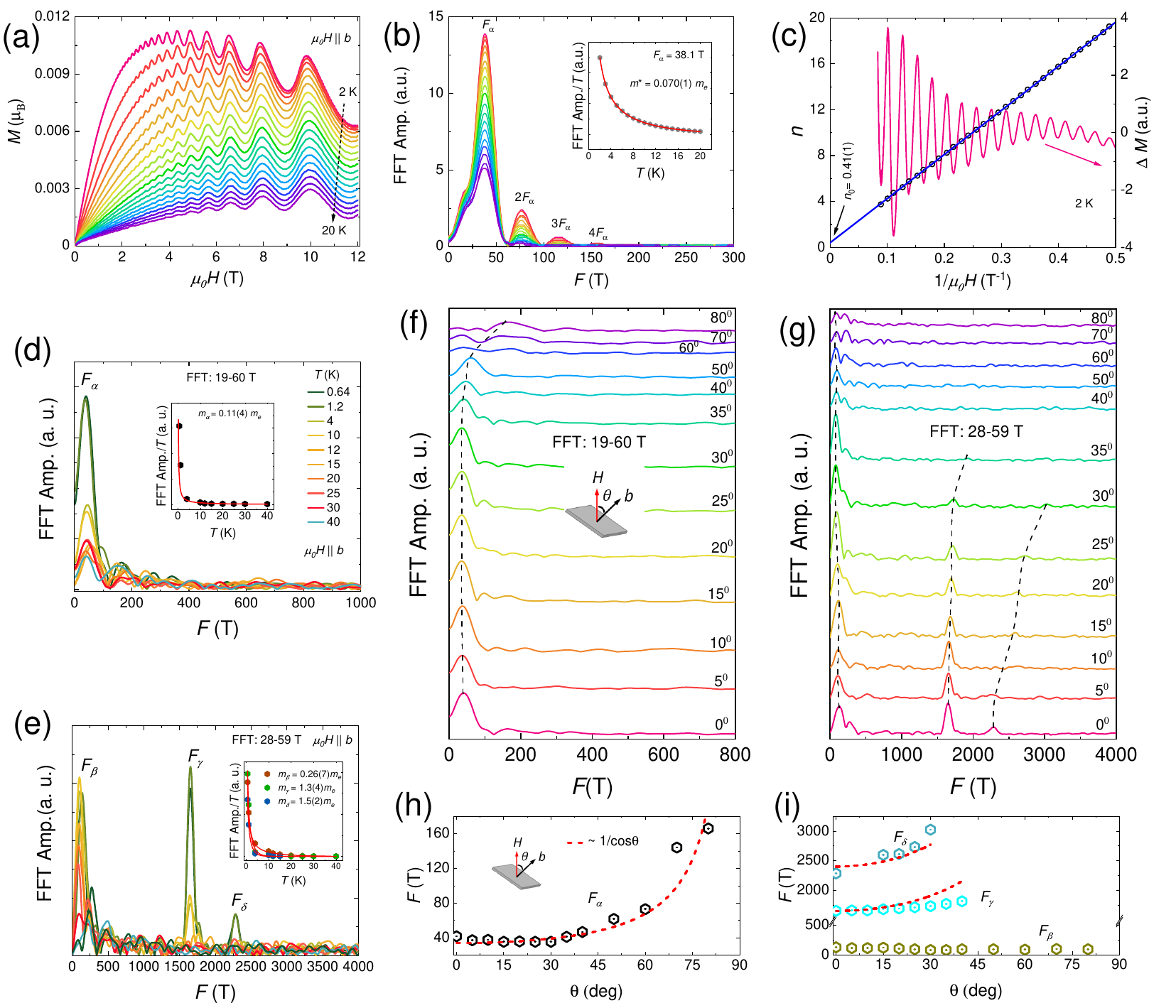}
	\caption{\label{QO} a)~Magnetization as a function of applied field for various temperatures measured along $H \parallel b$ up to 12 T. (b)~Fast Fourier transform spectra of dHvA-oscillation data for various temperatures. The inset shows the temperature dependence of the FFT amplitude fitted to the temperature-dependent part of the LK formula. (c)~Landau-level fan diagram along with the dHvA-oscillation data at 2~K. FFT spectra of SdH oscillations measured at 640~mK using the PDO for $H \parallel b$ for various temperatures for (e)$~19-60~$T and (f)$~28-59~$T field windows. Insets show the fitting of the FFT amplitudes to the temperature-dependent part of the LK equations.  (g)  and (h)~FFTs for two field windows for various angular positions with respect to the $b$-axis and applied magnetic field at 640~mK. (i) and (j)~Angular dependence of the oscillation frequencies. }
\end{figure*}
Field-dependent magnetization ($M$) measured with $H\parallel b$ for various temperatures exhibits strong de Haas-van Alphen (dHvA) quantum oscillations, as shown in Fig.~\ref{QO}(a). Such strong oscillations, even at high temperatures and low magnetic fields, suggest the high quality of our single crystals and a light effective mass. The corresponding fast Fourier transform (FFT) spectra for the field window 5.5-12~T are presented in Fig. \ref{QO}(b), indicating a fundamental frequency of $F_\alpha$ = 38.1~T along with its harmonics. The cyclotron effective mass ($m^*$) is estimated from the FFT amplitudes using the temperature-dependent part of the Lifshitz-Kosevich formula~\cite{Shoenberg}

\begin{equation}
    A_\text{FFT}/T = \dfrac{14.69m^*/B_\text{m}}{\sinh(14.69m^* T/B_\text{m})}
    \label{LK}
\end{equation}
where $B_\text{m}$ is defined as $1/B_\text{m} = \frac{1}{2}(1/B_\text{l} + 1/B_\text{u})$, with $B_\text{l}$ and $B_\text{u}$ the lower and upper field limits of the window, respectively. The inset of Fig.~\ref{QO}(b) shows the temperature-dependent FFT amplitude fitted to the above expression, yielding $m_\alpha ^*=0.070(1)m_\text{e}$. The observed oscillation frequency and its corresponding effective mass are very close to those in the previous report~\cite{YbBi2_2022}. To estimate the Berry phase ($\phi_\text{B}$), the Landau level (LL) fan diagram is constructed by assigning ($n$+1/4) and ($n$-1/4) to the maxima and minima of the dHvA oscillation measured at 2~K (Fig. \ref{QO}(c)). The Berry phase can be calculated from the intercept ($n_0$) of the linear fit to the LL fan diagram using the expression $\phi_\text{B}  = 2\pi (n_0  + \delta)$~\cite{PrAgBi2, LuSn2}. The phase-correction factor $\delta$ takes different values depending on the dimension of the Fermi surface. For a two-dimensional Fermi surface, $\delta$=0, whereas for a three-dimensional Fermi surface, it can be $\pm$1/8  for a minimum or maximum cross-section, respectively. Just as in the previous report and our Shubnikov-de Haas (SdH) quantum oscillation data described below, the field-orientation dependence  of the frequency $F_\alpha$ is two-dimensional in nature, suggesting $\delta$=0. Thus, the estimated $\phi_\text{B}  = 0.82\pi$, is nonzero, indicating nontrivial band topology in YbBi$_2$~\cite{YbBi2_2022,LuSn2}.

Further, the SdH quantum oscillations were measured at 640~mK using the proximity detector oscillator technique in a 60~T pulsed magnetic field. Field-dependent PDO signals and their corresponding oscillatory components at various temperatures and field orientations are presented in the Supplemental Material \cite{SM}. Fast Fourier transform (FFT) spectra of the oscillations measured at various temperatures are shown in Figs.~\ref{QO}(d) and \ref{QO}(e). Four fundamental frequencies are identified for $H||b$: $F_\alpha = 41.7$~T, $F_\beta = 127$~T, $F_\gamma = 1652$~T, and $F_\delta = 2286$~T. Two different magnetic-field windows, $19-60~$T and $28-59~$T, were used to obtain the lower and higher frequencies, respectively. 
Cyclotron effective masses were estimated by fitting the temperature dependence of the FFT amplitudes using the expression given in Eq.~\ref{LK}, as shown in the insets of Figs.~\ref{QO}(d) and \ref{QO}(e); the effective masses obtained for $H||b$ are $m_\alpha^* \approx 0.11m_\text{e}$, $m_\beta^* \approx 0.26m_\text{e}$, $m_\gamma^* \approx 1.3m_\text{e}$, and $m_\delta^* \approx 1.5m_\text{e}$. The effective mass associated with the $F_\alpha$ frequency is comparable to that obtained from the dHvA measurements and is consistent with the presence of relativistic fermions in YbBi$_2$~\cite{Liu2017}. Other Fermi-surface parameters, including the extremal cross-sectional area ($A$), Fermi wave vector ($k_\text{F}$), and Fermi velocity ($v_\text{F}$), for all the observed frequencies are summarized in SM~\cite{SM}. 
Figures~\ref{QO}(f) and \ref{QO}(g) show the FFT spectra obtained using two different magnetic-field windows while rotating the sample with respect to the applied magnetic field at 640~mK. The field-orientation dependence of the FFT frequencies is presented in Fig.~\ref{QO}(h) and \ref{QO}(i). The frequency $F_\alpha$ follows a $1/\cos\theta$ dependence, suggesting that the corresponding Fermi-surface section has a cylindrical shape. The frequency $F_\beta$ is nearly angle independent, whereas $F_\gamma$ and $F_\delta$ exhibit clear field-orientation dependences. The $F_\gamma$ frequency shows a weak angular dependence and deviates from the $1/\cos\theta$ behavior, indicating an anisotropic three-dimensional character. In contrast, the higher frequency $F_\delta$ follows a $1/\cos\theta$ dependence reasonably well over the angular range for which it is observed, suggesting that the corresponding Fermi-surface section has a quasi-two-dimensional character.

\subsection{ARPES and electronic band structure}

\begin{figure*}
\includegraphics[width=1.00\linewidth]{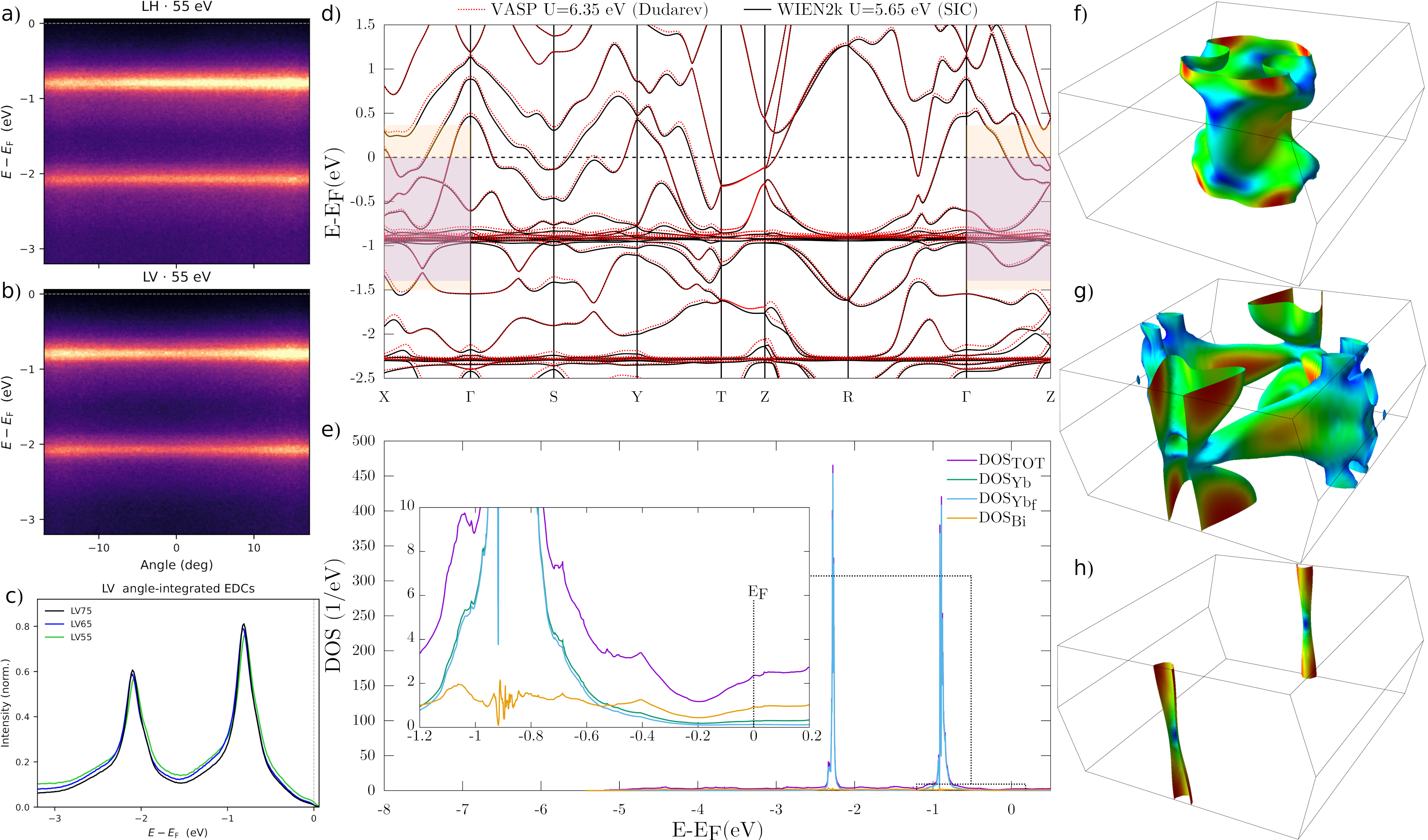}
    \caption{
    (a-c) ARPES (synchrotron) measurements of the  valence band. (a,b) maps show the photoemission intensity versus energy relative to the Fermi level ($E - E_{\rm F}$; $E_{\rm F} = 0$, white dashed line) and emission angle ($\pm 17^\circ$ about normal emission), recorded with linearly-horizontal (LH) and linearly-vertical (LV) polarized light at $h\nu$ = 55 eV.  Additional spectra for 65 and 75 eV are shown in Supplemental Material. Panel (c) show angle-integrated energy distribution curves (EDCs) for LV, with $h\nu$ = 75, 65 and 55 eV overlaid (black, blue, green). The spectra are governed by two essentially non-dispersive features at $E - E_{\rm F} \simeq -0.9$ and -2.3 eV originating from localized $f$ states;
    (d-h) Calculated characteristics of electronic properties of YbBi$_2$. d) Comparison of electronic dispersion relation calculated using {\sc wien2k} ($U=5.65$~eV) and {\sc vasp} ($U=6.35$~eV). Two distinctive flat bands correspond to localized, strongly correlated Yb$_{4f}$ states. Bands degeneration on T$\to$Z path yields importance for the topological properties of YbBi$_2$. We marked the areas of interest measured with ARPES (pink rectangles) and modeled with the DFT-based surface projection (orange). C.f., Fig. \ref{fig:vasp2arpes}. e) Density of states function for YbBi$_2$ valence band calculated in VASP. (f-h) The calculated Fermi surface composed of three bands (numbered 107, 109 and 111 respectively). The cylindrical shape of the third sheet of the Fermi surface is associated with the topological properties of YbBi$_2$ and its quasi-2-dimensional character.
    For the sake of completion, we refer the reader to the Supplementary Material, where we present the 3D and 2D (projected) Brillouin zones with the respective high-symmetry points.
    }
    \label{fig:electronic_structure}
\end{figure*}

Fig.~\ref{fig:electronic_structure} shows the ARPES data obtained from the synchrotron facility. Over the \(h\nu=55\text{--}75~\mathrm{eV}\) range, the valence-band region is dominated by the two flat, \(f\)-derived features, while the dispersing bands clearly resolved at low photon energy (\(h\nu=11~\mathrm{eV}\)) (discussed below) are absent. This photon-energy dependence is naturally explained by the combined action of photoemission matrix elements and the energy dependence of the photoelectron escape depth. 
First, at \(55\text{--}75~\mathrm{eV}\), the \(f\)-state photoionization cross-section is large compared with those of the \(s\)-, \(p\)-, and \(d\)-derived valence states, so the localized \(f\) emission dominates the measured spectral weight. Near \(h\nu=11~\mathrm{eV}\), the balance of cross-sections shifts in favor of the valence states, and the dispersing bands re-emerge. 
Second, \(55\text{--}75~\mathrm{eV}\) photoelectrons lie close to the minimum of the universal inelastic-mean-free-path curve (\(\ell \approx 5~{\AA}\)) and are therefore maximally surface sensitive. The associated out-of-plane momentum broadening, \(\Delta k_z \approx 1/\ell\), integrates over a large portion of the Brillouin zone along \(k_z\), smearing the dispersion of three-dimensional bulk bands while leaving the \(k\)-independent (dispersionless) \(f\) states unaffected. 
At \(h\nu=11~\mathrm{eV}\), the longer mean free path increases the bulk sensitivity and improves the \(k_z\) resolution, so that the bulk bands recover their dispersion. Because both mechanisms act in the same sense, together they account for the crossover from the \(f\)-dominated, non-dispersive spectra at \(55\text{--}75~\mathrm{eV}\) to the band-like spectra at \(11~\mathrm{eV}\). The persistence of the \(f\) features in both LH and LV polarizations is consistent with their localized, mixed-symmetry \(f\) character, whereas dispersing bands with well-defined orbital symmetry would be expected to show stronger polarization selectivity.

The sum of the LV and LH integrated spectra was then used to set the appropriate $U$ and $J$ values for the GGA+$U$ DFT calculations so that the computed density of states (DOS) reproduces the locations of the $4f_{5/2}$ and $4f_{7/2}$ peaks. Strong electronic correlations of the 4$f$ shell of Yb were required, as evidenced by the plain GGA DOS calculations, shown in Supplemental Material \cite{SM}. The electronic structure in the vicinity of the Fermi energy in this case is strongly dominated by Yb$_{4f}$ states, which exhibit a tendency toward localization; this leads to the formation of very narrow bands, manifested as sharp and narrow peaks in the density of states. The two $4f$ peaks, $4f_{5/2}$ and $4f_{7/2}$, are very close to the Fermi energy, much closer than the experimentally observed ones, showing that an additional Hamiltonian term derived from the Hubbard model~\cite{Hubbard} must be included in DFT calculations to account for on-site Coulomb interactions. The best match between the FP-LAPW method and the photoemission spectra was obtained by employing the double-counting self-interaction correction (SIC) scheme~\cite{SIC_Anisimov}, with the on-site Coulomb interaction parameter set to $U = 5.65$~eV and the exchange parameter set to $J = 0$~eV. The comparison of the density of states (convoluted with a Lorentzian with a full-witdth at half maximum, FWHM, value given in a figure) with the photoemission spectra is presented in Supplemental Material \cite{SM}. The calculated electronic dispersion relations, as shown in Fig. \ref{fig:electronic_structure}(a), exhibit two groups of narrow, flat bands originating from the Yb$_{4f}$ states.  Non-zero $J$ led to a larger, overestimated separation of the $4f_{5/2}$ and $4f_{7/2}$ peaks, which are split by the spin-orbit interaction. 
Noteworthy, applying the around-mean-field (AMF) double-counting scheme~\cite{AMF_Czyzyk} in the calculations induced a ferromagnetic state, in disagreement with the experimental result. 
To obtain consistent results with the experimental spectroscopy in the pseudopotential {\sc vasp} calculations (discussed below), a slightly higher value of $U = 6.35~$eV was necessary. This difference stems from the different computational schemes of the two methods, as {\sc vasp} uses the Dudarev GGA+$U$ approach \cite{Dudarev}.  After the position of $4f$ bands is adjusted, the remaining electronic structure from both all-electron and pseudopotential methods is the same, as shown in Fig.~\ref{fig:electronic_structure}.
\begin{figure*}
    \includegraphics[width=\linewidth]{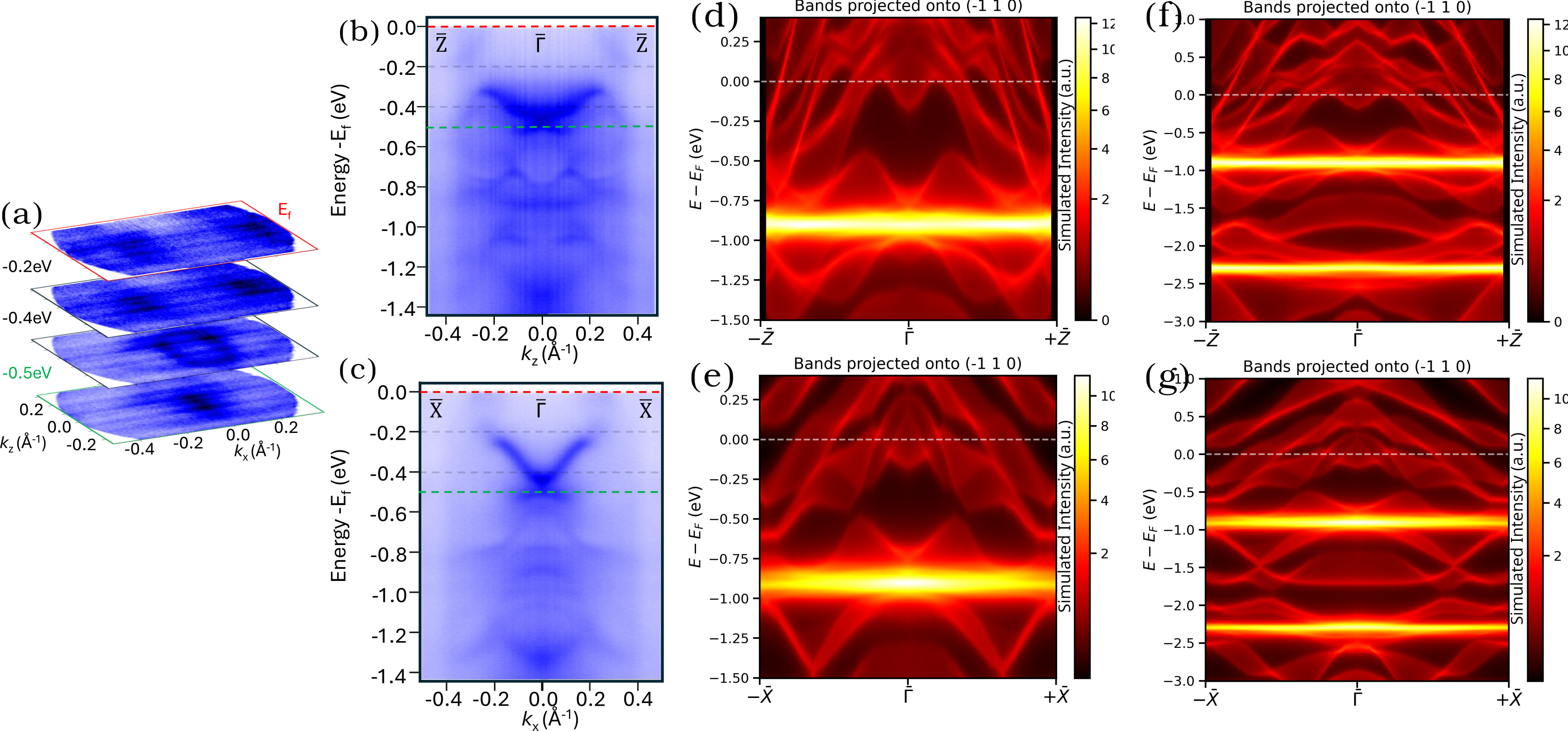}
    \caption{(a)-(c)~ARPES results.  Energy isosurfaces along $\bar{Z}$--$\bar{\Gamma}$--$\bar{Z}$ and $\bar{X}$--$\bar{\Gamma}$--$\bar{X}$ paths. Cuts across the electronic structure correspond to the iso-energy contours to the left.  Regions measured and recreated in DFT are marked in the band structure Fig. \ref{fig:electronic_structure}(a), while the high symmetry points in both full and reduced Brillouin zones are marked in Fig. \ref{fig:electronic_structure}(c). For the theoretical modeling, we project the bulk bands onto the (010) surface (elementary cell), equivalent to ($\bar{1}$10) in the primitive cell, of YbBi$_2$: in (d)-(e), we plot the projected band intensity near the Fermi energy $E_{\rm F}$ along the $\bar{\Gamma}\rightarrow\bar{Z}$ and $\bar{\Gamma}\rightarrow\bar{X}$ paths, respectively, while in (f)-(g) we present the projected band intensity for a broader energy range. In particular, the flat bands are predominantly displayed near $E_{\rm F}-0.9 \text{ eV}$ and $E_{\rm F}-2.3 \text{ eV}$.}
    \label{fig:vasp2arpes}
\end{figure*}

The density of states for the valence band near the Fermi level is presented in the inset of Fig.~\ref{fig:electronic_structure}(e). The total DOS at $E_\mathrm{F}$ is 1.290~(eV$^{-1}$) per formula unit (calculated with \textsc{wien2k}) and $N(E_\mathrm{F})=1.175$~eV$^{-1}$ per f.u. (calculated with \textsc{vasp}). Contributions to density of states at $E_{\rm{F}}$ projected onto atomic spheres and orbitals are summarized in Supplemental Material \cite{SM}. Using the experimentally determined Sommerfeld coefficient $\gamma = 4.8 \text{ mJ} \text{ mol}^{-1} \text{K}^{-2}$ the electron-phonon coupling constant $\lambda$  can be estimated using the renormalization of the bare band structure value $\gamma_{\rm band} = \frac{\pi^2}{3}k^2_\mathrm{B}N(E_{\rm F})$, $\lambda_{\rm ep} = \gamma/\gamma_{\rm band} - 1$ = 0.58 (calculated with \textsc{wien2k}) and $\lambda=0.73$ (calculated with \textsc{vasp}). These values generally confirm the weak-coupling regime of superconductivity in YbBi$_2$, but are slightly overestimated compared to $\lambda = 0.5$ estimated from the McMillan formula above. This may indicate some enhancement of the Coulomb pseudopotential, related to the vicinity of $4f$ electronic states to the Fermi level, as using the McMillan formula with a larger $\mu^* \sim 0.18$ would be required to obtain $\lambda \sim 0.6$.

We now return to the bulk electronic band structure for YbBi$_2$, shown in Fig. \ref{fig:electronic_structure}(d). Apart from the flat bands that span reciprocal space near $-0.9$~eV and deeper at $-2.3$~eV, a bulk Van Hove singularity is observed at the $\Gamma$ point between the electron-like $k_x/k_y$ band dispersions and their hole-like $k_z$ counterpart, located at $-0.6$~eV. 
Additionally, there is a fourfold band degeneracy along the $Z$--$T$ direction, indicating that these states are symmetry-protected. 
In order to reproduce the ARPES measurement, we used the outcome charge of the YbBi$_2$ structure to perform {\sc VASP} \cite{VASP} calculations for a dense mesh ($27 \times 27 \times 27 = 3^{9} = 19683$) k-points with symmetry optimization turned off (ISYM = -1). Using our original code~\cite{arpes_projector}, the bulk electronic states are projected onto the reduced (two-dimensional) Brillouin zone of the exposed surface (see Fig.~\ref{fig:electronic_structure}(c), which allows for a direct comparison with the ARPES data presented in Fig.~\ref{fig:vasp2arpes}, further validating our calculated band structure. 
Since the cleavage plane of single-crystal YbBi$_2$ is the $(010)$ plane, the reciprocal-space mappings in ARPES are also cut by the $x-z$ plane, making the in-plane momentum of the exposed surface lie along the $k_x$ and $k_z$ directions. The cuts along two perpendicular momentum directions $\Gamma -Z$, and $\Gamma - X$ are shown in Fig.~\ref{fig:vasp2arpes}(b) and Fig.~\ref{fig:vasp2arpes}(c), respectively. Along both directions, flat bands are visible at around $-0.85$~eV, indicating a very slight $p$-type Fermi-energy shift relative to theoretical modeling. Although the band dispersions and $4f$ states generally align with theoretical predictions for YbBi$_2$, the mappings reveal distinct differences in the $k$-dependence of crystal-field splitting within the Yb $4f$ states, manifested as intricate fine structure within the observed flat bands~\cite{vyalikh2010k}. 
The intensity of the flat bands at $-0.85$~eV is very low, likely due to matrix-element or surface effects, as these states are particularly susceptible to changes in the local electronic environment~\cite{neupane2013surface}. Near the bulk Van Hove singularity, there exists a Dirac-like dispersion. Such a feature is absent from the bulk band structure along the high-symmetry directions (c.f., Fig.~\ref{fig:electronic_structure}(a)) and exhibits the highest photoemission intensity of all observed bands. Its origin is clarified by the surface-projection calculations described above: when the bulk DFT states are projected onto the two-dimensional Brillouin zone of the (010) surface~\cite{arpes_projector}, the resulting spectral intensity along $\bar{\Gamma}-\bar{Z}$ and $\bar{\Gamma}-\bar{X}$ in Figs.~\ref{fig:vasp2arpes}(d)-(e) reproduces the Dirac-like dispersion observed in ARPES (c.f., Figs.~\ref{fig:vasp2arpes}(b)-(c)). The feature can therefore be attributed to surface-projected bulk states -- the envelope of bands dispersing weakly along the out-of-plane momentum $k_y$, reflecting the quasi-two-dimensional character of YbBi$_2$ -- rather than requiring a surface state localized by the broken translational symmetry at the cleaved surface~\cite{hagiwara2016surface}. The enhanced intensity of this feature may nevertheless signal an additional surface contribution, such as a surface resonance, which our projection-based approach, in contrast to slab calculations, cannot capture. Photon-energy-dependent ARPES measurements, probing the dispersion along $k_\perp$, would discriminate between the two- and three-dimensional character of these states~\cite{kealhofer2018observation}.

\subsection{Quantum oscillation calculations}

    

\begin{table}
    \centering
    \caption{Summary of the characteristic frequencies and effective masses obtained from experimental data and numerical calculations based on the Fermi-surface geometry. The indices $\alpha$, $\beta$, $\gamma$, and $\delta$ correspond to the frequencies identified in the experiments for $H||b$. $F_i \text{expt.}$ denotes the experimental value, $F_i \text{calc.}$ is the calculated value.}\label{tab:freq_m* comp}
    \begin{ruledtabular}
\begin{tabular}{l c c c}
& $f$ (T) & $m^* (m_\text{e})$ & band \\
\hline
$F_\alpha$ expt.   & 38.1 & 0.07 & --   \\
$F_\alpha$ calc. & 29   & 0.06 & 111   \\
$F_\beta$  expt.   & 127  & 0.26 & --   \\
$F_\beta$  calc. & 77   & 0.20 & 109   \\
$F_\gamma$ expt.   & 1652 & 1.3  & --    \\
$F_\gamma$ calc. & 2752 & 1.05 & 107    \\
$F_\delta$ expt.   & 2286 & 1.5  & --    \\
$F_\delta$ calc. & 3107 & 1.38 & 107    \\
\end{tabular}
\end{ruledtabular}
\end{table}

\begin{figure*}
    \centering
    \includegraphics[width=1.00\linewidth]{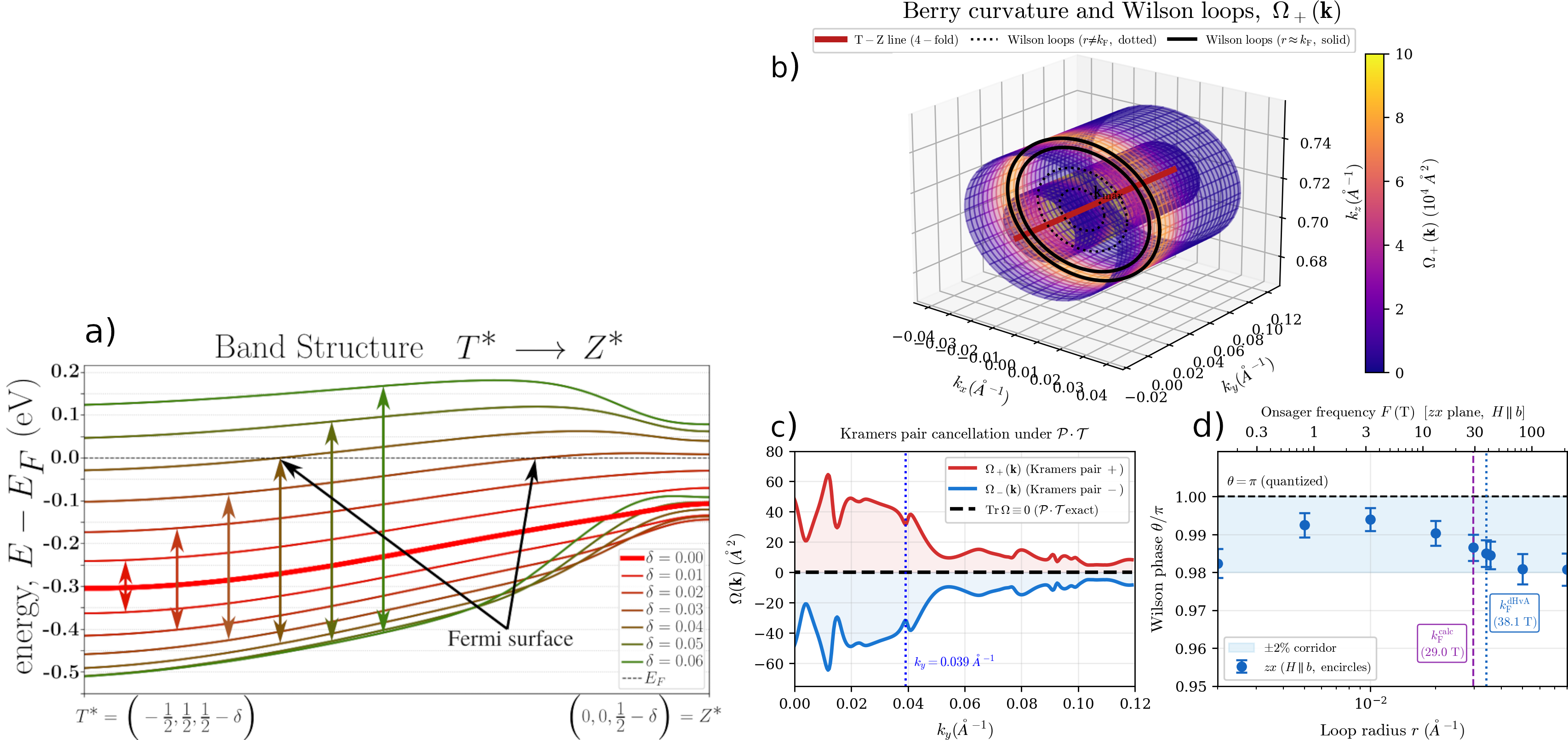}
    \caption{\textbf{a)} Split (colored arrows) of a degenerate T-Z band, leading to the crossing of the Fermi level (black arrow). \textbf{b)} Berry curvature and Wilson loops near T-Z line. The Fermi radii (both calculated and measured for $\alpha$ frequency) are marked with solid lines. \textbf{c)} Berry curvatures of a pair of bands and the center of the Fermi surface $\mathbf{k}_\text{max}\approx(0,0.039,0.710) (\AA{}^{-1})$ (marked with the blue dashed vertical line). Note that the curvatures cancel out exactly. \textbf{d)} Wilson phase $\theta = \sim\pi$ calculated around the $\mathbf{k}_\text{max}$ on the T-Z line. Onsager frequencies $F$ corresponding to the radii are marked on the top axis.
    }
    \label{fig:omega}
\end{figure*}

To numerically calculate the frequencies of quantum oscillations, the Fermi surface was computed and subsequently analyzed using {\sc skeaf}. The calculated Fermi surface, presented in Fig.~\ref {fig:electronic_structure}(d-f), consists of three distinct sections, as three bands cross the Fermi energy, as seen in Fig.~\ref{fig:electronic_structure}(d). Two tube-like FS sheets are observed in Fig. \ref{fig:electronic_structure}(f) and \ref{fig:electronic_structure}(h), reflecting the quasi-two-dimensional character of YbBi$_2$. The general shape of the FS is quite similar to that observed in CaBi$_2$~\cite{Golab2019}. 
For each experimentally observed frequency, the best-matching numerically calculated counterpart is listed in Table \ref{tab:freq_m* comp}.
Experimentally observed frequency $F_\alpha$ and small effective mass of the FS pocket which exhibits two-dimensional character, shows a closest match to the numerically calculated value for the band no 111, whose Fermi surface is shown in Fig.~\ref {fig:electronic_structure}(f). Higher frequencies differ quantitatively, with up to 20\% smaller calculated effective masses, compared to the measurement. 
The observation of a small effective mass from the dHvA and SdH oscillations measurements points to a Fermi-surface pocket associated with topologically nontrivial electronic states. This cylinder-like Fermi surface sheet is accurately reproduced by ab initio calculations and agrees well with experimentally determined frequencies and effective masses, making it suitable for further analysis of its topological properties.

\subsection{Topological properties}
\label{ssec:berri}
The analysis of the topological properties of the band structure is based on the {\sc vasp} results post-processed with the {\sc wannier90} code~\cite{Wannier90}. The four-fold degeneracy of the electronic bands near the Fermi level, along the $T\rightarrow Z$ path, is reduced to two-fold when we examine states near, but not on, that path (cf. Fig.~\ref{fig:omega}). For a small displacement $(\delta,\delta,0)$ in the fractional coordinates of the primitive cell — corresponding to a shift $\delta k_x$ along the Cartesian $k_x$ direction — the four-fold degenerate band splits into two branches (Kramers pairs), one above and one below the energy of the original band (vide Fig.~\ref{fig:omega}, left). To bridge the gap between DFT and real-space wavefunctions, we calculated the projections of the Bloch states onto a chemically intuitive basis set consisting of $s$-, $d$-, and $f$-orbitals for Yb and $p$-orbitals for Bi, ensuring a high-fidelity interpolation of the bands near the Fermi level.

As the crystalline structure is centrosymmetric and the compound is nonmagnetic, the space- ($\mathcal{P}$) and time-reversal ($\mathcal{T}$) symmetries of the crystal enforce the two-fold degenerate partner bands to have the Berry curvature cancel out, thus the calculation of Wilson phases (corresponding to each of the bands in a pair) is required with the Wilson loop integration being
\begin{align}
    \label{eq:wilsonLoop}
    W(\mathcal{C}) = \mathfrak{P} \exp \left( i \oint_\mathcal{C} A(\mathbf{k}) \cdot \mathrm{d}\mathbf{k} \right),
\end{align}
where $\mathcal{C}$ is the closed contour, $\mathfrak{P}$ the path-ordering operator, and $A$ the Berry connection. By choosing the loops in $zx$ plane (corresponding to the magnetic field being in the crystalline $b$ direction), with increasing radii, we can calculate the Wilson phases $\theta$ (phases of the $W(\mathcal{C})$ eigenvalues). Our result is within $\sim 2\%$ from the exact value of $\pm \pi$. The values of $\phi_B$ calculated from the experimental data and $\theta$ from the electronic band topology ($\phi_B\approx 0.82\pi$ and $\theta\approx 0.99\pi$ respectively) both should be seen as further indicators of the topological properties of YbBi\textsubscript{2}.

Regarding the nature of the topological properties, we utilize the {\sc IrRep} software suite\cite{irrep} to recover the wavefunction from the VASP output. This, in turn, allows for the parity analysis \cite{Fu20072} of the so-called \emph{ Time-Reversal Invariant Momenta} (TRIM), here all the possible permutations of fractional coordinates $n_i \in \{ 0, \tfrac{1}{2} \}$. While six of the TRIMs have odd parity ($(-1)^{11}$), both $\Gamma$ and $Y$ points have it even ($(-1)^{10}$ and $(-1)^{12}$, respectively). YbBi$_2$ exhibits a weak topological metal behavior with Fu-Kane-Mele \cite{Fu2007} indices $(0; 110)$. Moreover, the $Y$ point exhibits a small direct gap (46 meV). This proximity of the bands leads to the possibility that inducing a phase transition either by external or chemical pressure (e.g. doping with Pb) might cause a transition to a strong topological insulator, similarly to the one reported for ZrTe\textsubscript{5} \cite{Fan2017,Mutch2019,KovacsKrausz2024}.

\section{Conclusion}

In conclusion, we have performed a detailed study of the superconducting and topological electronic properties of high-quality YbBi$_2$ single crystals through thermodynamic, electrical transport, magnetic, quantum oscillation, and angle-resolved photoemission spectroscopy measurements, supported by first-principles calculations. Measurements of electrical transport, magnetization, and heat capacity on high-quality single crystals consistently demonstrate a type-I superconductivity near $T_{\mathrm{c}}\simeq 0.9$ K. The observed light effective mass indicates the presence of relativistic fermions in YbBi$_2$. The Fermi-surface topology observed from quantum oscillation and ARPES data is consistent with theoretical calculations, confirming the nontrivial character of YbBi$_2$. Our study establishes YbBi$_2$ as a rare example of a Yb-based superconductor in which superconductivity develops in the presence of topologically nontrivial electronic states. Our findings not only expand the family of Yb-based superconductors but also open a new route toward realizing topological superconductivity in rare-earth bismuthides.

\section{Acknowledgements}
This work was supported by the U.S. Department of Energy, Office of Science, Basic Energy Sciences, Materials Sciences and Engineering Division. The work at Gdansk University of Technology was supported by the Platinum Establishing Top-Class Research Teams project (DEC-2/2/2023/IDUB/I.1B/Pt). 
The work at AGH University was supported by the National Science Centre (Poland), Project No. 2025/59/B/ST3/01621.
We gratefully acknowledge the Polish high-performance computing infrastructure PLGrid (HPC Center: ACK Cyfronet AGH) for providing computer facilities and support within computational grants no. PLG/2024/017305 and PLG/2026/019296.  A portion of this work was performed at the National High Magnetic Field Laboratory, which is supported by the National Science Foundation Cooperative Agreement No. DMR-2128556, the US Department of Energy (DoE)  and the State of Florida. JS acknowledges support from the DoE BES FWP “Science of 100 T”.

\bibliography{references}

\end{document}